%% file: main.tex
\documentclass[10pt,conference]{IEEEtran}
\IEEEoverridecommandlockouts

\usepackage{cite}

\usepackage{amsmath,amssymb,amsfonts}
\usepackage{algorithmic}
\usepackage{graphicx}
\usepackage{textcomp}
\usepackage{xcolor}

\usepackage{multirow}
\usepackage{listings}
\usepackage{graphicx}
\usepackage{enumitem}
\usepackage{hyperref}
\usepackage{bbm}
\usepackage{booktabs}
\usepackage{tcolorbox}

\usepackage{tikz}
\usetikzlibrary{arrows,shapes, automata}
\usetikzlibrary{calc}
\usetikzlibrary{positioning}
\usetikzlibrary{arrows.meta}
\usetikzlibrary{fit}

\usepackage{pgfplots}

\pgfplotsset{
  compat=newest,
  xlabel near ticks,
  ylabel near ticks
}

\definecolor{testgreen}{HTML}{90EE90}
\definecolor{bfblue}{HTML}{AEE7F8}

\definecolor{javared}{rgb}{0.6,0,0} 
\definecolor{javagreen}{rgb}{0.25,0.5,0.35} 
\definecolor{javapurple}{rgb}{0.5,0,0.35} 
\definecolor{javadocblue}{rgb}{0.25,0.35,0.75} 
 
\DeclareMathOperator{\EX}{\mathbb{E}}

\newcommand{\tdot}[2][red,fill=red]{\tikz[baseline=-0.6ex]\draw[#1,radius=#2] (0,0) circle ;}

\newcommand{\Lagr}{\mathcal{L}}

\newcommand{\llbox}[1]{
	\begin{tcolorbox}[width=\columnwidth, colframe=black, boxrule=0.25mm, top=1mm, left=1mm, right=1mm, bottom=1mm]
		#1
	\end{tcolorbox}
}

\begin{document}

\title{DeepMutants: Training neural bug detectors with contextual mutations\thanks{This work 
has been partially funded by the German Research Council (DFG) under contract WE2290/13-1.}}

\author{\IEEEauthorblockN{Cedric Richter}
\IEEEauthorblockA{\textit{Carl von Ossietzky University Oldenburg} \\
Oldenburg, Germany \\
cedric.richter@uol.de}
\and
\IEEEauthorblockN{ Heike Wehrheim}
\IEEEauthorblockA{\textit{Carl von Ossietzky University Oldenburg} \\
Oldenburg, Germany \\
heike.wehrheim@uol.de}
}

\maketitle

\begin{abstract}
Learning-based bug detectors promise to find bugs in large code bases
by exploiting natural hints such as names of variables and functions or comments. 
Still, existing
techniques tend to underperform when presented with realistic bugs.
We believe bug detector learning to currently suffer from a lack of 
realistic {\em defective} training examples. In fact, real world bugs are scarce
which has driven existing methods to train on artificially created and mostly
unrealistic mutants.

In this work, we propose a novel {\em contextual} mutation operator which incorporates
knowledge about the mutation context to dynamically inject natural
and more realistic faults into code. Our approach employs a 
masked language model to produce a context-dependent distribution over feasible token 
replacements.
  The evaluation shows that sampling from a
language model does not only produce mutants which more accurately
represent real bugs but also lead to better performing bug detectors, 
both on artificial benchmarks and on real world source code.

\end{abstract}


\input{intro}

\input{basics}
\input{concept}

\input{evaluation}
\input{results}

\input{related}
\input{conclusion}

\section*{Acknowledgment}
The authors gratefully acknowledge the funding of this project by computing time provided by the Paderborn Center for Parallel Computing (PC²).

\bibliographystyle{IEEEtrans}
\bibliography{IEEEabrv,references}

\end{document}

%% file: intro.tex
\section{Introduction}
Learning-based approaches for bug detection have shown impressive results, 
often outperforming traditional methods in terms of accuracy and low false positive
rates \cite{habib2019neural, li2019improving, briem2019using, allamanis18graphs, vasic2019joint, hellendoorn2020great, pradel2018deepbugs, Ray2016Natural}. By exploiting natural hints embedded in code such as variable names or comments,
novel detection problems could be targeted that seemed to be infeasible before like finding misuses of variables \cite{hellendoorn2020great, vasic2019joint, allamanis18graphs},
detecting swapped arguments in function calls \cite{pradel2018deepbugs, karampatsis2020scelmo} or identifying one-off-errors \cite{briem2019using}.
Despite their visible improvement over traditional techniques and strong results on artificial benchmarks,
existing methods still show low 
performance when applied to real world code. 

We believe that the gap in generalization is not necessarily caused by the choice of learning algorithm,
but more by the lack of realistic bugs in training data. 
Although there exists an abundance of program code readily available in open source projects, bug-prone
code (i.e., code that reflects common coding mistakes of developers) makes it rarely into public codebases.
Data-driven methods, however, heavily rely on the availability of such code for learning accurate prediction models.  As a consequence, recent techniques in bug detection artificially increase the number of training examples
by injecting common bug types into correct code. While this type of data generation provides viable
training examples, the generation process itself is often based on pure randomness, ignoring the data distribution of realistic programmer mistakes and filling synthetic datasets with large quantities of atypical bug instances. Consequently,
the subsequent training risks overfitting the prediction model to simpler, unrealistic bug types. 

In this paper, we propose a novel approach for the generation of training data for neural bug detectors to overcome these current deficiencies. Our approach targets {\em single token bugs}.
Single token bugs represent the class of programmer mistakes which can be fixed by replacing a single token. This class includes well-studied examples  such as {\em variable misuses} \cite{hellendoorn2020great, vasic2019joint, allamanis18graphs}(a variable name is used although another was meant) and {\em one-off errors} \cite{briem2019using, pradel2018deepbugs} (the wrong use of a comparator in a boundary condition leads to an out-of-bounds error). 

Injecting realistic bugs is a long standing problem in software engineering, most prominently addressed in {\em mutation testing} \cite{Smith2009Augment, just2014major, namin2008sufficient, allamanis2016tailored}. Mutation testing targets the evaluation of test suites by mutating code to an erroneous state and
then running tests. Test suites detecting a large number of these so called {\em mutants} get a high evaluation because mutant detection has 
  been empirically shown to strongly correlate with real world bug detection \cite{just2014realbugs, daran96realbugs, papadakis2018mutation}.   Mutation testers produce and test mutants on the fly in large quantities. Mutation testing commonly relies on specific {\em stronger} mutations (certain injected bugs which are difficult to detect). 
The use of stronger mutants has been shown to improve both efficiency and effectiveness of the overall method \cite{just2012using, just2017inferring}. 

Our evaluation shows that the insights gained in mutation testing directly translate to the domain of neural bug detector training. In fact, bug detectors trained on dynamically generated strong mutants show a higher robustness and effectiveness on real world source code. 
Unfortunately, there does not exist a strong mutation operator for every possible bug type. 
To alleviate this issue, we propose a novel bug type independent {\em contextual} mutation operator, which itself {\em learns} to produce realistic mutants from data. More precisely, we continuously
train a masked language model \cite{devlin2019bert} together with a bug detector. The masked language model produces
a context-dependent probability distribution over all possible replacements for a masked token
with the objective of predicting the replaced token. Bugs are injected by masking out a random 
bug location and sampling from a restricted probability distribution produced by the language model.

Our operator is inspired by recent advances in natural language processing \cite{clark2020electra}, where it was shown that pre-training a generator that produces language mistakes trained together with a discriminator that detects introduced mistakes yields a superior language representation.
In this work, however, we are not interested in producing language representations but to generate 
realistic programmer mistakes. Therefore, we adapted the training method to our domain, including substantial
changes such as: (a) enabling training on correct unmutated program code, 
(b) stabilizing the training process by  a novel training scheduler
and (c) designing an efficient masking process to condition the mutation process
to a specific bug type. We implemented our method and tested it on several bug types in both Java and
Python software. Our experiments show that our proposed mutator produces more realistic mutants, which are substantially different from standard mutations. Overall, our training method improves the performance of bug detection on several
realistic as well as synthetic benchmarks, including a novel bug type for API misuses.

To summarize, we make the following contributions:
\begin{itemize}
\item we show that common strategies in mutation testing can be effectively transferred 
	to  the training of neural bug detectors,
\item we develop a novel bug-type independent contextual mutator, 
\item we experimentally show that contextual mutants and mutation learning can significantly improve neural bug detection. 
\end{itemize} 
We plan to make both implementation code and produced datasets available online.

%% file: basics.tex
\section{Foundations}
Our ultimate objective is the development of bug detection tools which can assist developers in improving code quality by finding potential bugs. We  start by discussing statistical bug detection in general 
and its underlying assumptions about source code.

\subsection{Statistical bug detection}\label{sec:stat-bug}
Developers tend to write code that helps in understanding and maintaining software systems \cite{allamanis2018survey}. As a consequence,
natural source code, that \emph{naturally} appears in software projects, shows far more regularities than we would expect
from an artificial language \cite{hindle2012natural}. These regularities can be exploited to build statistical models of code, which enable us to perform
tasks like code summarization \cite{alon2018code2seq, iyer2016summarizing, ahmad2020transformer}, code translation \cite{lachaux2020unsupervised} or statistical bug detection \cite{pradel2018deepbugs, arnar2020offside, Ray2016Natural}.
In the following, we focus on the latter task and give a brief overview over two approaches that target bug detection by finding statistical irregularities.

\textbf{Language models} exploit the fact that natural source code has similar statistical properties as natural languages \cite{hindle2012natural, karampatsis20bigcode}. These properties
include for example the re-occurrence of syntactical or semantic constructs~\cite{hindle2012natural, zhaopeng2014localness} and the co-occurrence of semantically related identifiers~\cite{allamanis2018survey}.
Based on this observation, language models, trained on large code corpora, model the probability of observing a program $C$ represented as a sequence of tokens $t_1, \dots, t_n$:
\begin{equation}
P(C) = P(t_1, \dots, t_n) = \prod_{i = 1}^n P(t_i \mid t_1, \dots t_{i-1})
\end{equation}
The last part of the equation is based on the observation that the joint probability of all tokens can be decomposed
in the conditional probability of observing $t_i$ given all previous tokens.\\
Interestingly enough, it was empirically shown that language models assign a lower probability to defective code, despite occurring naturally,
than to its correct counterpart~\cite{Ray2016Natural}.  As a result, language models were successfully applied in practice to identify
defective code lines and to support static bug finders~\cite{Ray2016Natural, karampatsis20bigcode, hellendoorn2017deep}.

\textbf{Neural bug detectors} address bug detection as a binary classification problem \cite{pradel2018deepbugs, arnar2020offside}. In contrast to language models,
they are explicitly trained to distinguish between defective and correct code. As a consequence, these bug classification models
require a potentially huge dataset of correct and incorrect code samples. While the first type of program code is easily obtainable e.g.~by scraping 
public code repositories, obtaining incorrect code is much more difficult. Defective code is far more seldom than correct code \cite{karampatsis2020sstubs}, 
defects are usually not annotated and manual annotations have the danger of enforcing an unwanted bias into the training process.
As a result, recent work started to seed artificial bugs into correct code to generate defective examples \cite{pradel2018deepbugs, arnar2020offside}. 
While the training on artificial bugs yields promising results on detecting real world bugs, these methods are often
limited by the bug types which they can produce.  For example, Pradel and Sen \cite{pradel2018deepbugs} built and trained well-performing models
for detecting arguments swaps, binary operator and operand replacements, but these models cannot be generalized to other bug types.

\textbf{Example:} To further motivate the use of statistical models for bug detection, we view the example 
depicted in Figure \ref{fig:null-pointer} taken from the Defects4J benchmark \cite{just2014d4j}. Note that the snippet is part of a method to retrieve information from a dataset.
In the second line, we query for a dataset object, which we use in the last line to obtain relevant information.
An experienced developer would expect that the \texttt{if}-statement performs a classical null-pointer-check and, hence, she would
recognize the software bug. Neither the compiler nor a classical static bug finder~\cite{spotbugs} could automatically identify the bug location. This is due to the fact that  
the code is correct in a different {\em context} and it does not break any coding conventions.
In contrast, a statistical bug detector could learn during training that an \texttt{if}-condition after a variable definition is likely a
null-pointer-check. Therefore, it assigns a high bug probability to the program snippet which ultimately guides
the developer to the bug location.

\begin{figure}
\centering
\begin{lstlisting}[language=Java]
int index = this.plot.getIndexOf(this);
CategoryDataset dataset = [...];
if (dataset (*@\colorbox{red!60}{!=}@*) null) { // ==
	return result;
}
int seriesCount = dataset.getRowCount();
\end{lstlisting}

\caption{Bug from Defects4J/Chart\#1}\label{fig:null-pointer}
\end{figure}

\begin{figure*}
\centering
\scalebox{0.9}{%
\input{images/mutation-framework}
}
\caption{Comparison between Mutation testing (\tdot[black, fill=testgreen]{2pt}) and the training procedure for a neural bug detection (\tdot[black, fill=bfblue]{2pt}). 
 }\label{fig:mut-comp}
\end{figure*}

\subsection{Neural Bug Detection vs.~Mutation Testing}\label{sec:nbf}
Bug seeding is not only an effective method to generate training examples in the context 
of neural bug detection. The value of artificial bugs have been extensively explored in the context of mutation testing \cite{Smith2009Augment, just2014major}.
Mutation testing approaches the qualitative evaluation of test suites by mutating existing code to produce artificial defective code. 
The produced code mutants often only differ in one randomly replaced operator or the removal of single statements. The effectiveness of a test suite
is then measured by a mutation score which is higher for test suites with tests failing on more mutants.

In the following, we give a more detailed overview over the mutation testing process\footnote{To further highlight the similarity between the two techniques, we describe mutation testing extended to augment test suites as proposed by Smith and Williams \cite{Smith2009Augment}. }  and we compare it to the training data generation used in neural bug detection~\cite{pradel2018deepbugs}.
Figure~\ref{fig:mut-comp} depicts the two processes with a simple example. Parts highlighted in green are specific to mutation testing and parts highlighted in blue are specific
to neural bug finding. The rest is shared between both processes.\\
We can identify the following three steps during both mutation testing and neural bug finding:
\begin{enumerate}[leftmargin=*]
\item{\emph{Generating mutants.}} A mutant is generated, here, by replacing a single operator in a likely correct code snippet (\texttt{<=} $\rightarrow$ \texttt{>}).
\item{\emph{Detection of mutants.}} Both real (unmutated) code and the generated mutants are fed to the detection system. A mutant is detected
by a test suite if at least one test fails during execution. The neural bug finder explicitly classifies the given code snippet, often 
without execution. If a detection system fails to distinguish a mutant from realistic code, then there is potential for improvement.
\item{\emph{Improve detection.}} Undetected mutants represent inaccuracies of a detection system, which can be 
mitigated in the context of mutation testing by adding new tests to a test suite. Neural bug finders, in contrast,
are optimized to learn from undetected mutants with the goal to improve their classification.
\end{enumerate}

The key difference between mutation testing and training neural bug detectors is the way mutants are generated: 
While a mutation tester produces new mutants {\em dynamically} at test time, neural bug detectors are typically
trained on a {\em static} set of mutants produced prior to training. In addition, mutation testers often employ
stronger mutation operators that guarantee compilable but harder to detect mutants \cite{just2012using, just2014major}.

%% file: images/mutation-framework.tex
\begin{tikzpicture}[node distance=3cm]

\node [rectangle, text width=2cm, align=center, minimum height=1.5cm, inner sep=2ex] (cb) {\textbf{Code Base} \\ \begin{lstlisting}[language=Java, xleftmargin=0pt]
if (x <= n)
						\end{lstlisting}};
						
\node [rectangle, text width=2.2cm, align=center] (mt) [above right=-0.5cm and 1cm of cb] {\textbf{Mutator} \\ \begin{lstlisting}[language=Java, xleftmargin=0pt]
if (x (*@\fbox{<=}@*) n)
						\end{lstlisting}};
						
\node [rectangle, text width=2.2cm, align=center] (mtt) [right=1.5cm of mt] {\textbf{Mutant} \\ \begin{lstlisting}[language=Java, xleftmargin=0.2cm]
if (x > n)
						\end{lstlisting}};
						
\node [rectangle, text width=2.2cm, align=center] (rl) [below=0.5cm of mtt] {\textbf{Real} \\ \begin{lstlisting}[language=Java, xleftmargin=0.1cm]
if (x <= n)
						\end{lstlisting}};

\node [rectangle, draw, inner sep=2ex, rectangle split,%
		rectangle split parts=2, rectangle split part fill={testgreen, bfblue}] %
(tb) [below right=-0.7cm and 1.5cm of mtt]
{

Test Suite

\nodepart{two}

Bug Detector

};

\node[draw,inner sep=2mm,fit=(tb), rounded corners ] (box) {};

\node (tba) [below=0.5cm of tb, minimum width=2.6cm] {\small (3) Update detection};

\node (mtq) [above right=-0.5cm and 1cm of tb]  {Mutant};
\node (rlq)   [below right=-0.5cm and 1cm of tb]   {Real};

\draw[-Latex, thick] (cb) edge node[above]{(1)} (mt.west)
	      (cb) edge node[below]{(2)} (rl.west)
	      (mt) edge node[above]{(2)}(mtt.west)
	      (mtt) edge node[above]{} (box)
	      (rl) edge node[below]{} (box)
	      (box) edge (mtq)
	      (box) edge (rlq);

\draw[thick] (tba) edge[-Latex, in=210, out=180] (box)
		(box) edge[-, out=-30, in=0] (tba);

\end{tikzpicture}

%% file: concept.tex
\section{Contextual mutants for bug detection}
We present a novel mutation framework for training neural bug detectors. Motivated by the observed
similarity to mutation testing, we aim to close the gap between the mutation procedure in mutation testing
and the training procedure in neural bug detection. In the process, we motivate and introduce a novel contextual
operator type, i.e., a mutation operator that incorporates the surrounding context for mutating program code.
During the rest of the paper, we focus on single-token bugs. 
Well-studied examples of this bug category are {\em OneOff}-errors \cite{briem2019using}, {\em binary operator replacements} \cite{pradel2018deepbugs}
and {\em variable misuses} \cite{allamanis18graphs, hellendoorn2020great, vasic2019joint}.

\subsection{Overview}
To train a learning-based bug detectors, as outlined in Section \ref{sec:nbf}, a mutation operator is required that transforms
likely correct code into likely incorrect code. Previous work~\cite{pradel2018deepbugs, hellendoorn2020great, vasic2019joint} addressed
this problem by hand-engineering specific mutation operators. The operator randomly selects a suitable bug location
where a single program token is replaced with a random alternative. While the operator likely produces a buggy program variant, 
relying on pure randomness might result in suboptimal unrealistic mutants \cite{just2017inferring}.
We consider the following example as instance where a pure random replacement leads to a suboptimal result.\\
\textbf{Example:} 
Figure \ref{fig:varmisuse} shows an example taken from the Defects4J benchmark \cite{just2014d4j}.
The original bug contained in the benchmark used the variable \texttt{p1} at the slot location, which can be fixed 
by replacing \texttt{p1} with \texttt{p2}. To produce a realistic mutant, an optimal mutation operator would have to reproduce
the original bug by also selecting \texttt{p1} as slot replacement. However, the operator employed in previous works \cite{hellendoorn2020great, vasic2019joint} 
selects a random variable in a definition or loading statement, such as \texttt{d1} or \texttt{it2}. As a result, the probability
of selecting \texttt{p1} and, hence, producing a realistic mutant is less than 15\%. In contrast, a contextual operator, 
which is conditioned on the surrounding context, would assign \texttt{p1} a higher sampling probability
(e.g.~by detecting that \texttt{p1}  is the only type correct replacement).

\begin{figure}
\centering
\begin{lstlisting}[language=Java, xleftmargin=0cm]
PathIterator it1 = p1.getPathIterator(null);            
PathIterator it2 = (*@\colorbox{red!60}{<S>}@*).getPathIterator(null);             
double[] d1 = new double[6];            
double[] d2 = new double[6];            
boolean done = it1.isDone() && it2.isDone();
\end{lstlisting}

\caption{VarMisuse Bug based on Defects4J/Chart\#11.}\label{fig:varmisuse}
\end{figure}

This observation has motivated the development of a {\em contextual mutation operator}. Our approach is inspired by recent advances
for pre-training language models in natural language processing~\cite{clark2020electra}. More precisely, it was shown
that training a generator that produces token replacements jointly with a discriminator that detects introduced
replacements yields superior language representations. 
Here, we apply this idea in the domain of bug detection. 
This domain shift has lead to substantial changes in the pre-training task
which we present throughout the rest of this section. 

\begin{figure*}
\centering
\scalebox{1.0}{%
\input{images/framework}
}
\caption{Joint training framework for contextual mutations and bug detectors.}\label{fig:arch}
\end{figure*}
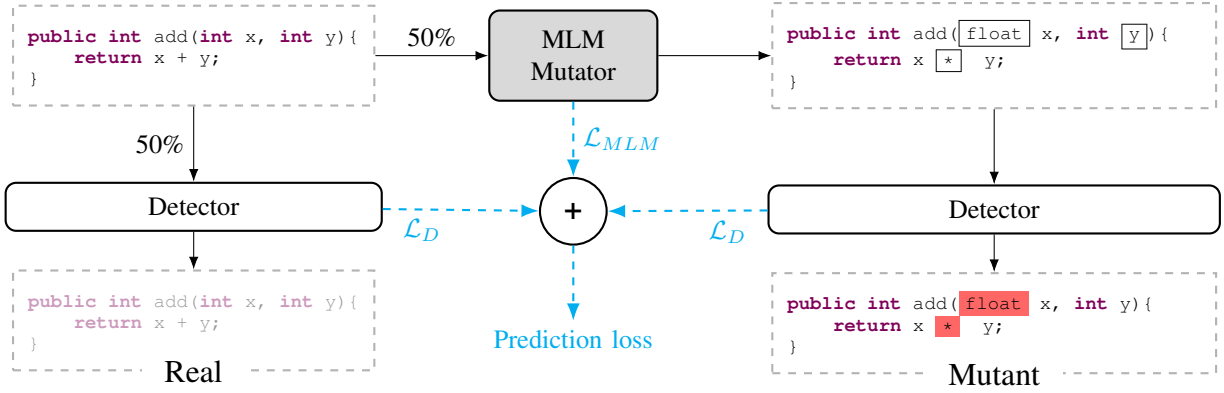

Our mutation framework for training neural bug detectors is depicted in Figure~\ref{fig:arch}. To produce
a balanced distribution of buggy and non-buggy training examples, we feed 50\% of all code snippets to the mutator
and 50\% are directly processed by the bug detector. After sampling random bug locations, our mutator
predicts a replacement distribution using a masked language model (MLM). The detector is then tasked
to classify unmutated code as ``Real'', mutated code as ``Mutant'' and has to identify error locations in mutated code. We jointly
optimize the detection capabilities of the bug detector and the likelihood of the mutator for producing
realistic replacements.

\subsection{Program encoding}

Throughout the rest of the paper, program code is encoded as a sequence of tokens $[CLS], t_1, t_2, \dots, t_n, [EOS]$. The tokenization closely 
follows the programming language syntax by parsing and traversing the program AST\footnote{Our work employs \texttt{tree-sitter}, an AST parser that supports a plethora of programming languages: \url{tree-sitter.github.io}}. $[CLS]$ is a special token which will allow us to obtain classifications over the whole sequence.

Following Karampatsis et al. \cite{karampatsis20bigcode}, tokens are split into subtokens via a byte-pair encoding (BPE) \cite{sennrich2016neural}.
We employ the best performing BPE with at least 10k merges. Note that the subtoken splitting is mostly
used to decrease the computational footprint and to handle novel unseen identifiers during testing. All our employed
mutators will operate on token level. 

To explore the effect of our framework in isolation, we follow previous work by employing a sequence-based neural language encoder that maps a sequence 
of input tokens $\mathbf{x} = [x_1, x_2, \dots, x_n]$ to a contextualized vector representation $[h_1, h_2, \dots, h_n]$. 
If not stated otherwise, we utilize a BERT-like transformer architecture \cite{devlin2019bert}. Note however that the general framework
is model-agnostic and both mutator and bug detector can be replaced by more sophisticated neural encoding models 
like graph neural networks \cite{li2019improving} or AST-based models \cite{hellendoorn2020great}.

\subsection{From static to dynamic mutant generation}
The biggest difficulty of producing mutants during training time is the computational footprint of the respective
mutation operator. In fact, some of the existing operators require AST-based static analysis to perform mutations.
Consider, for example, the injection task in Figure \ref{fig:varmisuse}. To produce a VarMisuse bug, 
local variable declaration and variable usages have to be identified for finding an injection point and a potential replacement. 
We noticed, however, that in virtually all cases the heavy lifting appears during the selection of mutation targets and
the search for replacement candidates, while applying traditional operators is a cheap random process.
As a consequence, we perform the computational intensive part in the preprocessing step
by annotating the program with positions of mutation targets and a list of replacement targets.
Replacement targets are either program locations or program tokens like operators, function names or language keywords. 
Finally, these annotations enable us to apply mutation operator on-the-fly during the training process.

\subsection{Masked language mutants}
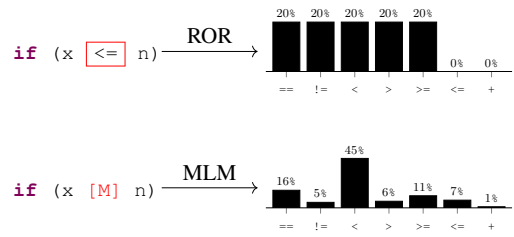
\begin{figure}
\centering
\scalebox{0.85}{%
\input{images/mlm-mutation}
}
\caption{Comparison of relational operator replacement (ROR) with our MLM mutator.}\label{fig:mlm-mut}
\end{figure}

Next, we describe our contextual mutation operator as a sampling process on a masked language model (MLM).
We start by giving a general intuition for MLMs in the context of programming languages.\\
\textbf{Masked language models} (MLM) \cite{devlin2019bert} extend classical language models with the objective
to reproduce masked out tokens. The following code snippet depicts a typical task for a MLM:
\begin{lstlisting}[language=Java, xleftmargin=0.2\columnwidth]
for(int i = 0; i (*@\color{red}{[M]}@*) n; i++)
\end{lstlisting}
The goal is to predict the original comparator (\texttt{<}) at the masked location based on the given context.
MLMs produce a probability distribution over all possible replacements assigning higher probability
to tokens that occur more frequently in a similar context during training (\texttt{<=}, \texttt{!=}, \texttt{>}). To further extend our intuition,
we trained an MLM on a corpus of Java code and visualized the predicted sampling distribution in 
Figure \ref{fig:mlm-mut}. We can observe that the MLM assigns a high probability to the same
tokens, which would be sampled by a traditional relational operator replacement (ROR) mutator,
while giving a higher probability to replacements that are more likely in this context.
Notice, for example, that the MLM ranks relational operator differently than we would expect in our for-loop example, because
the input to the MLM is an If-statement instead.
\textbf{Replacement mutations} modify source code by randomly replacing single operators or literals, as we have seen in our previous example.
Formally, we can define this behavior as two distinct sampling operations on a program $T = t_0 \dots t_n$:
\begin{align*}
	m \sim P(\text{type}(t_i) = \color{green!60!black}{\checkmark} \color{black}) && \hat{t}_m \sim P( r \mid t_m, C),
\end{align*}
where the first sampling operation uniformly samples a mutation position $m$ from the set of program tokens that qualify for replacement ($\text{type}(t_i) = \color{green!60!black}{\checkmark} \color{black}$).
Afterwards, a replacement $\hat{t}_m$ is sampled from the distribution of all replacements $r$, which is conditioned on the current token $t_m$ and a potential further context $C$.
As an example, Just et al. \cite{just2017inferring} proposed to restrict the set of possible replacements conditioned on the surrounding abstract syntax tree. 

Our proposed \textbf{masked language mutation} substitutes the replacement distribution by an MLM:
\begin{equation*}
	P( r \mid t_m = \color{red}\text{[M]}\color{black}, T_{\text{masked}}),
\end{equation*}
where $T_{\text{masked}}$ is the program $T$ we want to mutate, except that a program token $t_m$ selected for mutation
is replaced by a special mask token \color{red}[M]\color{black}. A key advantage of using MLMs instead of traditional replacement distributions is that the probability
distribution is generally defined for all token types. As a result, an MLM mutator is able to perform language {\em specific} replacement, e.g.~replacing variable types
in typed languages or modifying function calls.\\
\textbf{MLM training: } The MLM sampling distribution is produced by a neural language encoder
with the training objective of predicting the original masked out token $t_m^{org}$. We can address this training
objective by minimizing the expected log-likelihood of predicting the original token over all sampling operations \cite{devlin2019bert}:
\begin{equation*}
\Lagr_{MLM} = \EX \left[  -\log P( t_m^{org} \mid t_m = \color{red}\text{[M]}\color{black}, T_{\text{masked}}) \right]
\end{equation*} 
Instead of pre-training an MLM mutator, we follow Clark et al.~\cite{clark2020electra} and train
our mutator continuously together with the bug detector. Note that we can view this process 
as a form of adaptive mutant generation. In fact, at the beginning of training, tokens are nearly randomly replaced, giving rise to a higher number of easy bugs 
for the bug detector. As the training progresses, the probability of sampling context-dependent replacements increases, giving rise
to far more difficult error types. Finally, since the risk of resampling the original token also increases during training, we sample from a
restricted probability distribution by setting the probability of the original token to zero and renormalizing the probability of the remaining tokens. 

\subsection{Neural bug detection \& localization}\label{sec:bugdetect}
Given a code snippet as token sequence $T = [t_0, t_1, \dots, t_n]$ either generated by our mutator or coming directly from a dataset, 
a second neural encoder produces a buggyness score $b_i$ per token $t_i$. The buggyness score is a real valued scalar, 
where a higher score is assigned to a program location containing a bug with a higher likelihood. To jointly
perform localization of errors and classification of program instances, we employ a version 
of the pointer network proposed by Vasic et al.~\cite{vasic2019joint} restricted to localization and classification.\\
\textbf{Bug pointer network:} Assuming there is only one bug location in the program or at least 
the program can be fixed by iteratively repairing single bug locations, we can apply
a pointer architecture that models the probability of a token being buggy as a softmax distribution:
\begin{equation*}
P(t_i \text{ buggy}\mid T) = \frac{\text{exp}(b_i)}{ \sum_{j \in M} \text{exp}(b_j)},
\end{equation*}
where $M$ is a set of indices for potential bug locations.
A token $t_i$ has a higher bug probability if it achieves a higher bugginess score $b_i$ relative to the other tokens in the program.
The token $t_0$ (the [CLS] token) has the special function of representing correctness of a program snippet. In other words, if 
the code snippet is correct, we have a high probability at the $[CLS]$ token, otherwise a high probability should be assigned to the bug location. 
The goal is achieved by minimizing the expected negative log-likelihood over all error locations for mutated programs and [CLS] locations for realistic programs:
\begin{equation*}
\Lagr_D = \EX \left[  - \log P(t_i \text{ buggy} \mid T) \right]
\end{equation*}

Finally to train both mutator and bug detector, we minimize the combined loss:
\begin{equation*}
\Lagr = \Lagr_{MLM} + \lambda \Lagr_{D},
\end{equation*}
where $\lambda$ is a hyperparameter tuned such that $\Lagr_{MLM}$ and $\Lagr_{D}$ have the same impact on the overall loss.

\subsection{Technical details}
In this section, we present technical contributions which helped to stabilize 
the overall optimization process and enabled the training on a single consumer grade GPU.\\
\textbf{Mutation pipelining:} Neural networks are typically trained on batches of training 
examples randomly drawn from a dataset. Simulating this batching behavior
in our mutation framework, 
by mutation a random number of training examples per batch, would mean that the mutator is trained on fewer training examples than the detector,
and the batch size of the mutator has a high variance resulting in a high variance of the mutator loss function.
As a result, the optimization of both mutator and detector diverges, which prohibits learning.

\begin{figure}
\centering
\includegraphics[width=0.50\textwidth]{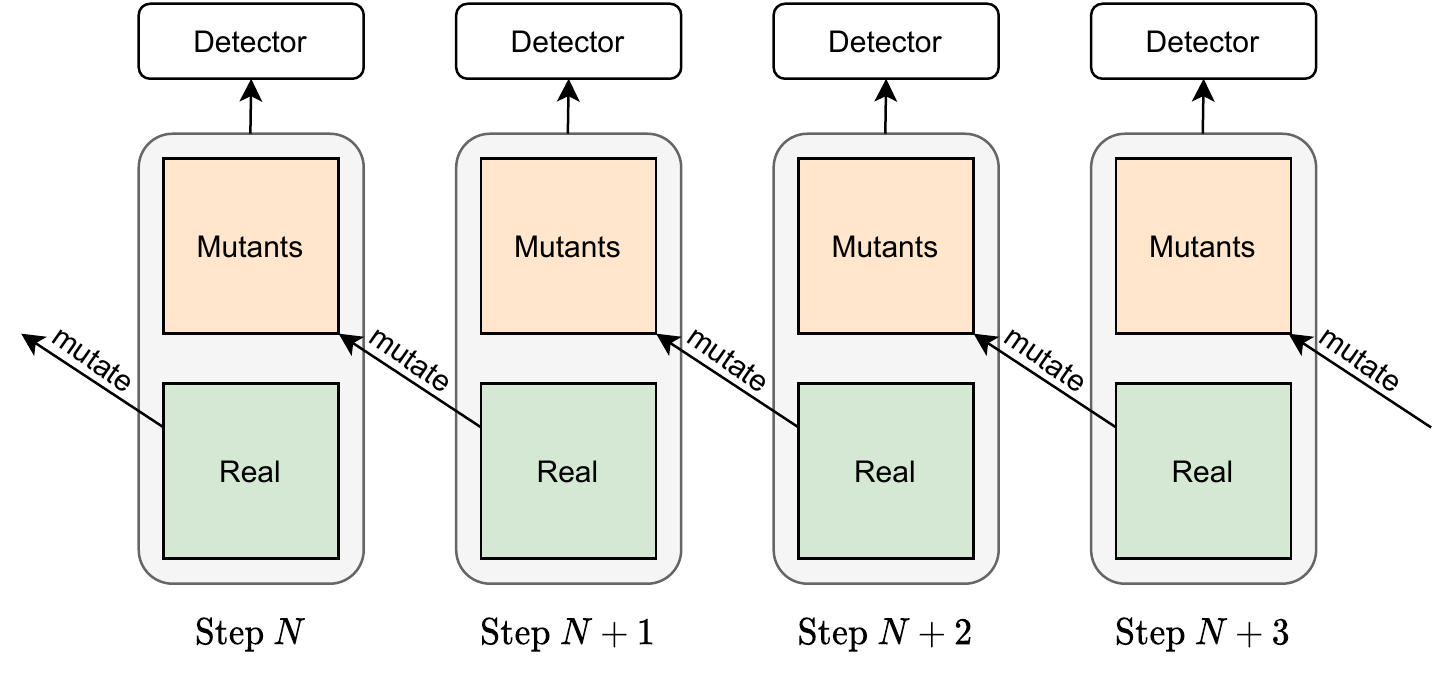}
\caption{Mutation pipelining.}\label{fig:mut-pipe}
\end{figure}

To mitigate this effect, we propose to pipeline the mutation as shown in Figure \ref{fig:mut-pipe}. 
Now, the training happens on two independent batches, one consisting of realistic examples and the other consisting of mutated examples. 
After one training step, the batch of mutated code is discarded, while the batch of realistic code is given 
to the mutator for producing new mutants for the next step. As a result, both detector and mutator are 
trained on fixed batch sizes, while both models observe all training examples during training.\\
\textbf{Random offset augmentation:}
The employed Transfomer architecture uses absolute position information. In other words,
the Transformer knows that the token $t_n$ is at the $n$th position, which could lead
to overfitting the Transformer to bugs at specific program locations. Previous work \cite{shaw2018relative} has therefore explored
relative position information to mitigate the effect. However, these techniques create a substantial computational 
overhead. Instead, we propose to augment the absolute positions with a random offset. Since, in the view of the Transformer, the positions of program tokens are constantly moving, overfitting
to specific positions is harder and the tasks can only be solved by learning relative position information.\\
\textbf{Approximate Length Batching:}
Training the neural language encoder with sequence-level batching (i.e., by fixing the number of training sequences per batch)
inflicts a high GPU memory consumption. When training with function code on a single GPU with 11GB memory,
we are restricted to a maximum of 128 functions per batch with a length of maximally 128 subtokens. Increasing
the sequence length has in most of our experiments a quadratic effect on the batch size and, hence, increases the training time more than quadratically.
For this reason, we follow Hellendoorn et al.~\cite{hellendoorn2020great} and employ a token-level batching of 12,5K tokens per batch with functions
restricted to a length of 250 subtokens. We found that close to 90\% of all functions in our datasets are below the 250 subtokens threshold. 
To further increase the training efficiency, functions of similar length are batched together. Note that, although the number of functions per batch
has a high variance, the number of tokens per batch is fixed and, hence, we did not observe training instabilities 
when combined with mutation pipelining.

%% file: images/framework.tex
\begin{tikzpicture}

\node [dashed, thick, draw=gray!60, inner sep=0.2cm] (real) {\scalebox{0.8}{\input{images/addfnreal}}};

\node [thick, draw, fill=gray!30, rounded corners, text width=2cm, align=center, minimum height=1.2cm](mutator) [right=1.5cm of real] {MLM \\ Mutator};

\node [dashed, thick, draw=gray!60, inner sep=0.2cm, text width=5.5cm] (mut) [right=1.5cm of mutator] {\scalebox{0.8}{\input{images/addfnmut}}};

\node [thick, draw, rounded corners, inner sep=0.2cm, minimum width=5cm] (detect1) [below=1cm of real] {Detector};
\node [thick, draw, rounded corners, inner sep=0.2cm, , minimum width=6cm] (detect2) [below=1cm of mut] {Detector};

\node [dashed, thick, draw=gray!60, inner sep=0.2cm, text opacity=.4] (reall) [below=0.5cm of detect1] {\scalebox{0.8}{\input{images/addfnover}}};
\node [dashed, thick, draw=gray!60, inner sep=0.2cm, text width=5.5cm] (mutl) [below=0.5cm of detect2] {\scalebox{0.8}{\input{images/addfnlabel}}};

\node[thick, draw, circle, inner sep = 0.2cm] (add) [below=1cm of mutator] {\large \textbf{+}};
\node (loss) [below=1cm of add] {\color{cyan}Prediction loss};

\node[fill=white, inner sep=0.3cm] (reallabel) [below=-0.45cm of reall] {\large Real};
\node[fill=white, inner sep=0.3cm] (mutlabel) [below=-0.45cm of mutl] {\large Mutant};

\draw[-Latex] (real) edge node[above]{50\%} (mutator)
		(mutator) edge (mut);
		
\draw[-Latex] (real) edge node[left]{50\%} (detect1)
		(detect1) edge (reall);
		
\draw[-Latex] (mut) edge (detect2)
		(detect2) edge (mutl);
		
\draw[-Latex, cyan, dashed, thick] (mutator) edge node[right]{\color{cyan}$\Lagr_{MLM}$} (add)
					     (detect1) edge node[below, near start]{\color{cyan}$\Lagr_{D}$} (add)
					     (detect2) edge node[below, near start]{\color{cyan}$\Lagr_{D}$} (add)
					     (add) edge (loss);
\end{tikzpicture}

%% file: images/mlm-mutation.tex
\begin{tikzpicture}

\node [rectangle, text width=2.2cm, align=center] (mt) {\begin{lstlisting}[language=Java, xleftmargin=0pt]
if (x (*@\fcolorbox{red}{white}{<=}@*) n)
						\end{lstlisting}};
						
\node [rectangle, text width=2.2cm, align=center] (mlm) [below=1cm of mt] {\begin{lstlisting}[language=Java, xleftmargin=0pt]
if (x (*@\color{red}{[M]}@*) n)
						\end{lstlisting}};

\node (uni) [right=1.5cm of mt] {\scalebox{0.6}{\input{images/uniform-dist}}};

\draw[->] (mt) edge node[above] {ROR} (uni);

\node (mlmd) [right=1.5cm of mlm] {\scalebox{0.6}{\input{images/mlm-dist}}};
\draw[->] (mlm) edge node[above] {MLM} (mlmd);

\end{tikzpicture}

%% file: evaluation.tex
\section{Evaluation}
Our evaluation investigates the effect of the mutation operator used to produce artificial bugs on the 
performance of a bug detector. In the process, we study not only the choice of mutation operator (contextual or predefined),
but also the bug injection mode (static or dynamic). Our exploration aims to answer the following research questions: 
\begin{description}
\item[RQ1] How does the mutation operator affect the evaluation and real world performance of a neural bug detector?
\item[RQ2] Does the bug injection mode influence the effectiveness of a bug detector?
\item[RQ3] Do contextual mutations produce more realistic software bugs?
\end{description}
We designed individual experiments to answer our research questions. In the following, we start by describing the datasets,
metrics and baselines used in our evaluation.

\subsection{Datasets}
During our evaluation, we will employ varying code corpora for training and evaluating bug detectors. In the following, we describe them in detail. \\
\textbf{Unsupervised code corpus:}
Motivated by the availability of official bug benchmarks in Java such as Defects4J \cite{just2014d4j} and ManySStubs4J \cite{karampatsis2020sstubs},
we focus our evaluation on bugs in Java methods. As a corpus of likely correct Java code, we choose the public portion
of the CodeSearchNet challenge \cite{husain2019codesearch}. While the corpus contains examples for multiple languages,
we utilize the Java subset containing 500K Java methods divided into train, validate and test split at ratio of 80-10-10.
All methods are deduplicated both in and across splits \cite{husain2019codesearch}. 

As results on a code corpus of Java methods can only be representative for programming languages
that are strongly-typed and compilable, our evaluation includes experiments for a second set of Python functions.
In this scenario, we utilized the widely adopted ETH Py150 \cite{ray2016model} dataset containing 150K Python files crawled from Github.
More specifically, we employed the deduplicated and preprocessed version proposed by Hellendoorn et al.~\cite{hellendoorn2020great} and use the same split
into 90K training files, 10K files for validation and a 50K files test set. The final Python datasets consist of the extracted
top-level function found in the program code. 
During our experiments, we constructed several dataset versions where we vary the type of injected bug  
and the frequency of injection. We plan to publish the preprocessed datasets for future comparisons.\\ 
\textbf{Real world bug benchmark:}
In addition to testing on artificial bugs, testing our models on real world bugs is part of our investigation.
Therefore, we choose the publicly available defect benchmark ManySStuBs4J \cite{karampatsis2020sstubs}. ManySStubs4J is a benchmark
automatically mined from over 100 Github projects with a focus on single line bugs. The dataset divides over 10K bugs in 16 bug categories, in which
more than 80\% of classified bugs are replacement bugs. For each of our bug type, we selected the corresponding bug category
in the benchmark. Bugs that either do not occur in a method body or are contained in a method with more than 250 tokens are excluded. 

\subsection{Bug types}\label{sec:btypes}
To generate training and evaluation datasets, we populate the unsupervised code corpora with artificial bugs. In this study,
we focus on three different bug categories. In the following, we give an overview of the studied bug types:
\begin{description}[leftmargin=0cm]
\item{\textbf{Binary operator replacement (BOR)}} addresses bugs related to binary operators. Typical examples
are OneOff-errors but other real world examples such as the null-pointer check in Figure~\ref{fig:null-pointer} belong
to the same category. We haven chosen BOR bugs as a category studied in the context of several previous methods 
\cite{pradel2018deepbugs, briem2019using, karampatsis2020scelmo} and with the availability
of elaborated mutation operators found in the mutation testing literature \cite{just2014major}. In particular,
we seed bugs by arithmetic (AOR), relational (ROR), conditional (COR) and bitwise (BIT) operator 
replacements. An example of the ROR operator can be found in Figure~\ref{fig:mlm-mut}.
In addition, we experimented with a {\em weaker} mutator variant employed by previous bug detectors
which just samples a random binary operator replacement from the set of all binary operators.
\item{\textbf{Variable misuse bugs (VarMisuse)}} describe a bug category where a local variable 
is replaced by another variable defined in the same context. These types of bugs often occur
when code is replicated for another use case, while variables are not replaced by mistake.
An example is depicted in Figure \ref{fig:varmisuse}. 
Similar mutant categories do not exist in mutation testing. Therefore, we employed
the data generation method proposed by previous works \cite{vasic2019joint}, which gives us the opportunity
of testing the effectiveness of our contextual mutants in comparison.
\item{\textbf{API misuse bugs (APIMisuse)}} is a bug category which we
propose to study a broader range of bugs. APIMisuse bugs occur when a
programmer chooses a wrong function during a function call. A common pitfall
for developers are typically functions with a similar name or signature \cite{karampatsis2020sstubs}.
As a simple mutator baseline, we collect a vocabulary of function calls
from our training corpus. To mutate a function call,
we sample from the vocabulary and the set of occurring function calls.
For example, the following function call can be mutated as follows:
\begin{lstlisting}[language=Java, xleftmargin=0.15\columnwidth]
graph.hasNodes() -> graph.hasEdges()
\end{lstlisting}
\end{description}
\textbf{Metrics:} During our evaluation, we report {\em Classification accuracy} (whether the detector identifies
a bug in a function) and {\em Localization accuracy} (whether the model correctly identifies 
the bug location in a buggy examples). The later metric is only reported when applicable.

\subsection{Neural bug detectors}
To study the effect of the training dataset on generalization performance of neural bug detectors, 
we train several baseline tools for neural bug detection.\\
\textbf{Deep word vectors} is a simplified variant of the learning-based
bug detector DeepBugs~\cite{pradel2018deepbugs} for binary operator replacement bugs.
We utilize the same extraction heuristics as DeepBugs to produce triplets
consisting of operator, left and right operand (AST-based information are ignored). 
The triplets are embedded to a learnable vector representation and then fed to a neural network classifier.
We do not pre-train embeddings and use the same hyperparameters as proposed by the original
authors.\\
\textbf{BiLSTM} acts as an alternative for a sequence-based neural language encoder based on
bidirectional LSTMs \cite{hochreiter1997long}. Based on the results of Hellendoorn et al. \cite{hellendoorn2020great}, we employ
the best performing model architecture with two layers and a hidden dimension of 512.
For bug detection, we use the pointer network inspired by Vasic et al.~\cite{vasic2019joint} and described in Section \ref{sec:bugdetect}.
We tuned learning rate, dropout rate and weight decay to avoid overfitting per bug category and experiment.\\
\textbf{Transformer} is a 6-layer BERT-like transformer architecture  with a hidden dimension of 512, 8 attention heads,
a dropout rate of $0.1$. The transformer employs the same pointer model as the BiLSTM baseline.
We found that a learning rate between $5e-5$ and $2e-4$ generally works best for 
all our tasks. Additionally, we experimented with sinusoid positional embeddings but found that learned embeddings
together with our random offset augmentation worked better. During our contextual mutation experiments,
we additionally use a 4 times smaller Transformer to generate mutations.\\
\textbf{Training:}
In general, we found that a learning rate scheduler improved the results for all neural language encoders. We therefore 
warmup the learning rate for 10K steps and decay afterwards. All our models except the Deep word vectors
are trained with an Adam optimizer \cite{kingma2015adam}. To achieve comparable results with Deep word vectors,
we use the same optimizer as used by DeepBugs. All neural language encoders are trained
for at least 24h on a single GPU for the respective bug detection task.\\
\textbf{Additional baselines:}
During our main experiments, we do not include graph-based neural architectures like GNNs \cite{allamanis18graphs} or the newly introduced GREAT model \cite{hellendoorn2020great}.
These models usually use a graph representation that is task- and language-dependent, which cannot be easily generalized to other bug types.
For similar reasons and since our main focus is bug detection, we do not  compare with more complex detect and repair models like Hoppity \cite{dinella2020hoppity}.

%% file: results.tex
\section{Results}

\subsection{RQ1: How does the mutation operator affect the evaluation and real world performance of a neural bug detector?}
We start our investigation with an experiment to confirm our intuition that pure random mutants 
are insufficient for training neural bug detectors. We are disregarding contextual mutants for RQ1. \\
\textbf{Experimental design:} In this experiment, we explore two different mutator variants
for injecting binary operator replacement bugs. The first mutator which we name weak mutator
produces buggy examples by sampling a pure random binary operator. This is equivalent
to the data generation process used by recent methods~\cite{pradel2018deepbugs,briem2019using}.
The second mutator called strong mutator is based on the state of the art
in mutation testing for generating binary operator bugs~\cite{just2014major}, which we described in Section~\ref{sec:btypes}.

As a baseline bug detector, we employ the deep word vector model and train two variants
based on the used mutation type called \texttt{Deep-W} (weak mutation) and \texttt{Deep-S} (strong mutation).
To produce training and test set, we generate a balanced dataset of over 2M examples from our Java corpus for training
and over 60K example for validation by extracting examples and mutating binary operators.
In total, we produced two versions each for training and test set by swapping the mutation operator.
The set based on weak mutation is called \texttt{Mut-W} and the set based on strong mutations is called 
\texttt{Mut-S}. We evaluate across datasets and report classification accuracy.
As a real world test set, we extracted 284 examples\footnote{Examples are excluded with unknown identifier because in this case both models can only guess.} of real world binary operator bugs and their fix from the ManySSubs4J benchmark.\\
\textbf{Result:} Our experimental results are reported in Table \ref{tab:abugs}. All measurements are reported in 
classification accuracy over bug candidates. Best results are highlighted in bold. While comparing 
the performance of the same detection model trained and evaluated on different BOR bug distributions, we 
make the following observations:
\begin{description}
\item[Bug detectors perform best on mutants they are trained on]
The trained bug detectors perform better on the respective mutant types they are trained
on, while significantly underperforming on the respective other mutant type. 
This suggests that the bug detector, as a statistical model, fits the mutant distribution generated by the mutator.
\item[Stronger mutations lead to stronger detectors]
Training on strong mutations outperforms the training on weak mutations on real world bugs (with a gain
of 5.82\%). In addition, ranking detectors with respect to their performance on strong mutants is a better
proxy for real world performance.
\end{description}
The results suggests that the used mutation operator is not only important for training, but 
also has a strong effect on the evaluation of bug detectors. While real world tests could
help to achieve more representative results, the choice of mutator could have
a significant impact on the development process\footnote{Large scale evaluation on artificial bugs often guide the development of novel bug detection systems.} and model selection of a bug detector.
\llbox{
In summary, a bug detector trained on stronger mutants
can achieve a stronger performance on real bugs. The evaluation on strong mutants
can be a better proxy for the real world performance.
}
\begin{table}
\caption{Classification accuracy for bug detection on weak and strong mutants}\label{tab:abugs}
\begin{center}
\resizebox{0.65\columnwidth}{!}{%
\begin{tabular}{lccc}
\toprule
& \multicolumn{3}{c}{\textbf{Test-Set}} \\
\cmidrule{2-4}
\textbf{Model} & \textbf{Mut-W} & \textbf{Mut-S} & \textbf{Real} \\
\midrule
Deep-W & \textbf{89.02\%} & 66.51\% & 53.88\% \\
Deep-S &70.19\% & \textbf{73.39\%} & \textbf{59.70\%}\\
\bottomrule
\end{tabular}%
}
\end{center}
\end{table}

\subsection{RQ2: Does the bug injection mode influence the effectiveness of a bug detector?}
\begin{table}
\caption{The effect of data scaling and contextual mutations on bug detection performance}\label{tab:dynamic}
\begin{center}
\resizebox{\columnwidth}{!}{%
\begin{tabular}{lccccccccc}
\toprule
\multirow{3}{*}{\textbf{Model}} &%
\multicolumn{6}{c}{\textbf{Java}} & &%
\multicolumn{2}{c}{\textbf{Python}} \\
\cmidrule{2-7} \cmidrule{9-10}
& \multicolumn{2}{c}{\textbf{BOR}} & \multicolumn{2}{c}{\textbf{VarMisuse}}  & \multicolumn{2}{c}{\textbf{APIMisuse}} && \multicolumn{2}{c}{\textbf{VarMisuse}}  \\  
\cmidrule{2-7} \cmidrule{9-10}
& Cls & Loc & Cls & Loc & Cls & Loc && Cls & Loc \\
\midrule
\textbf{Static(1x)} & & &&&& && &  \\
\hspace{0.3cm}BiLSTM  & \textbf{92.07} & 88.43 & 71.85 & 49.03 & 72.33 &  52.24 & & 63.43 & 48.14 \\
\hspace{0.3cm}Transformer & 90.24 & 85.85 & 89.95 & 81.17 & 71.12 & 49.70 && 82.26 & 73.44 \\
\textbf{Static(3x)} & & &&&& && &  \\
\hspace{0.3cm}BiLSTM  & 91.92& 91.15& 76.78 & 64.71 & 75.35& 62.09 & & 76.57 &  64.72 \\
\hspace{0.3cm}Transformer & 91.25 & 89.85 & 91.21 & 84.45 & 72.02 & 51.16 && \textbf{90.27} & 79.71\\
\textbf{Dynamic} & & &&&& && &  \\
\hspace{0.3cm}BiLSTM & 91.96 & \textbf{91.54} & 82.18& 74.13 & \textbf{78.22} & \textbf{72.05} && 80.18 & 75.88 \\
\hspace{0.3cm}Transformer & 91.38 & 88.76& \textbf{91.87} & \textbf{85.09} & 73.26 & 52.19 && 89.64 & \textbf{81.87}\\
\midrule
\textbf{Contextual} & & &&&& && &  \\
\hspace{0.3cm}CM-0  & 89.44 & 87.04 & 76.32 & 48.51 & 72.77 & 53.83 & & 84.51 & 71.79 \\
\hspace{0.3cm}CM-250K  & 92.49 & 90.85 & 91.39 & 83.12& 84.26 & 72.38 && 91.05 & 83.10 \\
\hspace{0.3cm}CM-2M & \textbf{92.54} & \textbf{91.66} & \textbf{93.02} & \textbf{85.12} & \textbf{85.35} & \textbf{78.38} &  & \textbf{91.85} & \textbf{86.39} \\
\bottomrule
\end{tabular}%
}
\end{center}
\end{table}
One of the main differences between mutation testing and training bug detectors is the mutation frequency.
Therefore, we study the effect of moving from static to dynamically generated datasets.\\
\textbf{Experimental design:} Now, we consider all three introduced bug variants
by employing the different mutators on the respective datasets. For binary operators, we employ
the stronger mutation operator based on our results in RQ2. There exists no stronger mutator version
for the other bug types. Hence, we employ the standard mutator version described in Section \ref{sec:btypes}.
To answer RQ2, we construct three different datasets varying in size per task and per programming language:
\begin{description}[leftmargin=0cm]
\item[\textbf{Static (1x)}] is constructed by mutating functions exactly once. The dataset includes
one buggy and one correct version for each example in our code corpus.
\item[\textbf{Static (3x)}] is constructed by creating multiple mutants per function. The number of mutants 
is restricted to maximal three distinct mutants. Each mutant is paired with a corrected version and included in our dataset. 
\item[\textbf{Dynamic}] does not include any mutants at all. Mutated examples are constructed on-the-fly during training.
\end{description}

For evaluation, we require a static dataset to achieve comparable results across experiments. For this reason,
we generate one evaluation set per bug type using the method described in Static (3x). For VarMisuse on Python
functions, we use the validation set provided by Hellendoorn et al. \cite{hellendoorn2020great}.

During our experiments, we compare the validation performance of different neural language encoder
trained on the constructed datasets. We do not explicitly report the results for the deep word vector model on BOR,
since it was already shown  that they perform worse than neural encoder on that task.\\
\textbf{Result:} The results of our experiments are shown in Table \ref{tab:dynamic}. 

\begin{figure}
\centering
\includegraphics[width=0.35\textwidth]{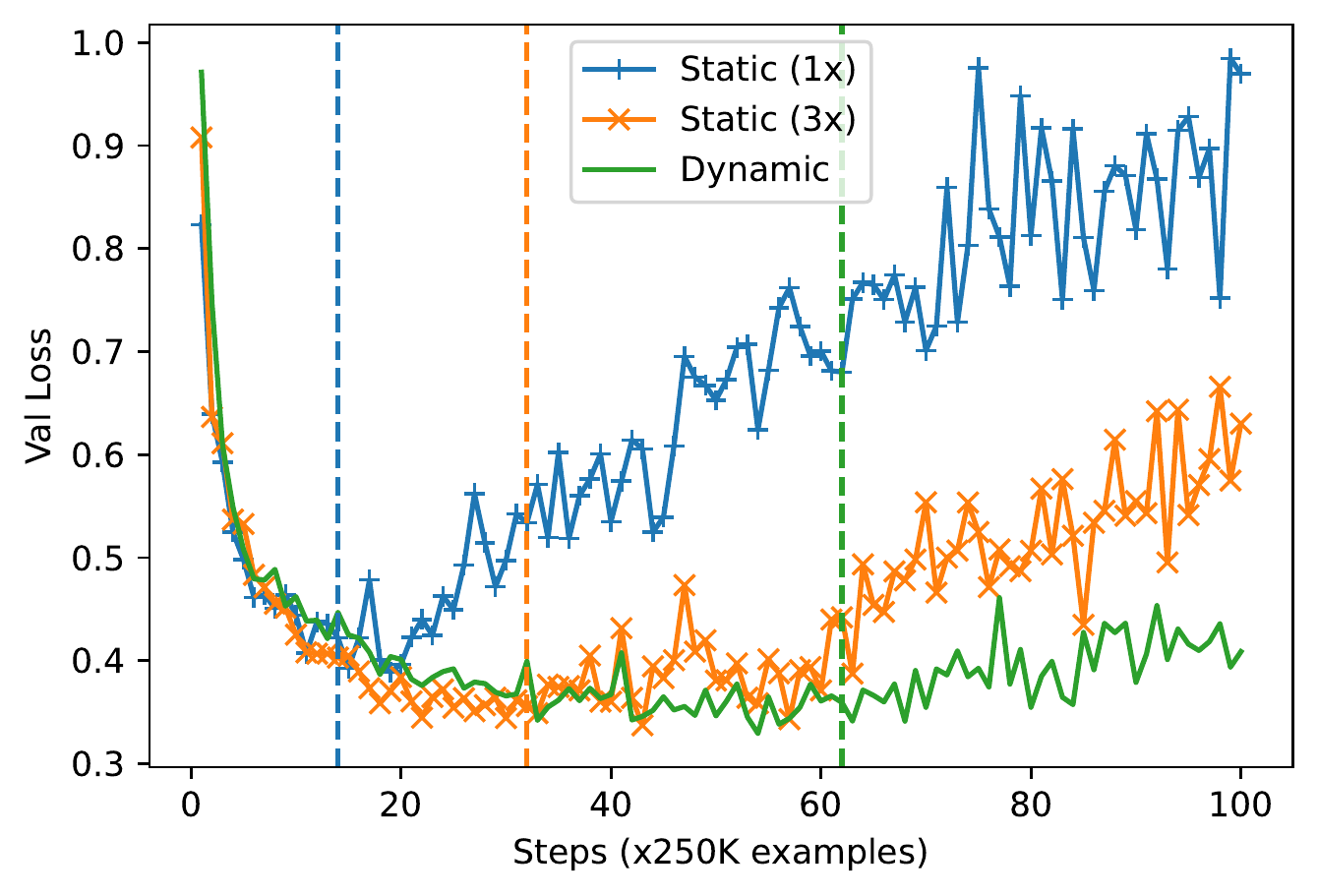}
\caption{Validation loss of Transformer with different mutator but same hyperparameter on BOR. Vertical lines mark point of overfitting.}\label{fig:val_loss}
\end{figure}

We report the maximal validation accuracy achieved per model in terms of classification and localization accuracy.
We refrain from reporting results on a test set due to the concerns raised in RQ1. Instead, 
we directly report real world performance of the best performing models in Table \ref{tab:real}.
The rows for contextual mutations can be ignored for now.
We make the following observation for the validation experiments:
\begin{description}
\item[\textbf{More mutants help more}:]
In general, we see that more data increases the validation performance
on all bug types and detector types, sometimes even achieving
gains of up to 15.68\% in bug localization performance and up to 5.4\%
by only increasing the number of mutants per function.
\item[\textbf{Dynamic mutants never hurt bug detection}:]
Switching from 3 times mutation to dynamic mutants never decreases
bug localization performance. While in most cases, the classification
accuracy increases, there exists examples where the classification
accuracy decreases in favor of an improvement in bug localization.
\end{description}
In addition to the aforementioned results, we found that some models 
overfit on the static BOR datasets after some time, while dynamic
mutations act as a strong regularizes by delaying or stopping overfitting.
We provide an example for the validation loss of an Transformer trained on 
the BOR dataset in Figure \ref{fig:val_loss}. All models share the same hyperparameter
in this example, but we trained on different dataset versions.
\llbox{
In summary, a bug detector can fit the mutator distribution better when mutations
are generated dynamically. In addition, dynamic mutations have a regularizing
effect on the detector.
}

\subsection{RQ3: Do contextual mutations produce more realistic software bugs?}
\begin{table}
\caption{Real world performance on ManySStubs4J}\label{tab:real}
\begin{center}
\resizebox{\columnwidth}{!}{%
\begin{tabular}{lcccccc}
\toprule
\multirow{2}{*}{\textbf{Model}} & \multicolumn{2}{c}{\textbf{BOR(171)}} & \multicolumn{2}{c}{\textbf{VarMisuse(118)}}  & \multicolumn{2}{c}{\textbf{APIMisuse(839)}}   \\  
\cmidrule{2-7}
& Cls & Loc & Cls & Loc & Cls & Loc \\
\midrule
Static (BiLSTM) & 59.31 & 22.22 & 52.68 & 10.38 & 51.72 & 6.38\\
Static (Transformer) & 61.51 & 34.44 & 61.05 & 21.45 & 51.30 & 4.22 \\
Dynamic (BiLSTM)& 63.76 & 29.26 & 53.69 & 12.72 & 51.87 & \textbf{10.86} \\
Dynamic (Transformer) & 57.15 & 35.56 & 58.38 & 18.46 & 51.03  & 3.85 \\
\midrule
CM-0 (Transformer) & 62.04 & 43.33 & \textbf{71.13} & \textbf{32.53} & 50.02 & 8.72\\
CM-2M (Transformer) & \textbf{63.81} & \textbf{48.15}& 63.07 & 24.48  & \textbf{52.78} & 9.64\\
\bottomrule
\end{tabular}%
}
\end{center}
\end{table}
After studying the effect of common mutation testing techniques on bug detection performance,
we investigate if contextual mutation can further improve the results.\\
\textbf{Experimental design:} 
We train one bug detector per bug type using the dataset of the dynamic setting
but employ the proposed contextual mutation instead of a fixed mutator. 
As a detector, the same Transformer architecture is used as in previous experiments with the same hyperparameter. 
We utilize a 4 times smaller Transformer as a mutator. Since the generator will likely
not produce the same bug distribution we evaluate on, we additionally continue 
the training of the discriminator in the dynamic setting. Overall, we report on
3 versions: CM-0, CM-250K and CM-2M. CM-0 is only trained on contextual mutants,
while CM-250K and CM-2M is further trained on 250K and 2M examples 
produced by a fixed mutator, respectively.\\
\textbf{Results:} Our results on the validation set for all three models are reported in Table \ref{tab:dynamic}
and real world performance for the best performing detectors is reported
in Table \ref{tab:real}. For the real world experiments, we additionally report the number of examples
per test set. Overall, we make the observation that: 
\begin{description}
\item[\textbf{Contextual mutants improve real world effectiveness}]
While training on pure contextual mutants (CM-0) significantly underperforms
on all synthetic benchmarks, this does not replicate to the real world. Sometimes
training on a fixed mutator after contextual mutations (CM-2M in Table \ref{tab:real})
decreases the detection performance.
\item[\textbf{Tuning towards synthetic bugs improves results}]
Further tuning the bug detector to the validation distribution for one epoch (CM-2M)
leads to a bug detector beating all baselines on the synthetic datasets. Sometimes
it is enough to train the detector for 1/8 of an epoch. However, as expected,
this does not always improve results on real world tasks. On VarMisuse
the bug localization performance decreases by 8.05\%, although the 
performance on the synthetic task has increased by nearly 36.61\% after tuning.
\end{description}
Overall, we find that training on contextual mutants improves the bug localization
performance both on synthetic and real world benchmarks when considering
the same model architecture. The strong performance
of the Transformer solely trained on contextual mutants (CM-0) is a strong 
indicator that the produced bug distribution is more realistic.
\llbox{
In summary, contextual mutants can provide gains 
for a learning based system both on synthetic tasks
and on real world benchmarks.
}

%% file: related.tex
\section{Related Work}
In this work, we explored a novel data generation method for training neural bug detectors
on artificial program bugs. We next discuss how our techniques relate to previous work
in statistical bug detection, contextual mutations for mutation testing and general techniques
for training data generation in language processing.

{\em Machine learning and statistical bug detection.}
As outlined in Section \ref{sec:stat-bug}, there exists two main research
directions in statistical bug detection based on language modeling~\cite{karampatsis20bigcode, Ray2016Natural, hellendoorn2017deep}
and explicit bug classification~\cite{pradel2018deepbugs, allamanis18graphs, vasic2019joint, hellendoorn2020great}. Language modeling addresses 
bug detection mostly in an unsupervised way by modeling
the statistical likelihood of program code and viewing bugs 
as statistical irregularities~\cite{karampatsis20bigcode, Ray2016Natural, hellendoorn2017deep}. 
Ray et al. \cite{Ray2016Natural} provided evidence that buggy code is in fact more irregular
than correct code in the view of an ngram language model. Follow-up work \cite{karampatsis20bigcode}
showed that neural network based language models are even better suited for this task.
In this work, we address a contrary problem by employing a masked language model
to introduce more natural bugs into programs. By exploiting statistical regularities to 
produce bugs, our bug instances are more difficult for unsupervised language models while 
providing a strong learning signal to supervised bug classifier. Supervised bug classifiers
treat bug detection as a classification task by learning from both correct and erroneous code.
DeepBugs \cite{pradel2018deepbugs} learned to identify bugs in binary operators or functions
by mining positive training examples from a corpus of JavaScript code and mutating them to produce buggy 
examples. Allamanis et al. \cite{allamanis18graphs} introduced a wide range of naming bugs
into C\# code (including VarMisuse bugs) and employed a graph based neural network
to identify them. Vasic et al. \cite{vasic2019joint} identified and repaired introduced VarMisuse bugs
in Python functions with a novel pointer network. Hellendoorn et al. \cite{hellendoorn2020great}
showed that the performance of pointer networks can be improved by switching 
the employed neural network type. While it seems that the main source for
advances in neural bug detection is the invention of new neural program
representations, we showed that the type of data generation also has high
impact on the performance of neural detection systems. More precisely, all our bug detectors
use  neural architecture similar to the one explored by Hellendoorn et al.~\cite{hellendoorn2020great},
while we report significant performance gains both on realistic tasks and synthetic benchmarks
by improving on the bug injection strategy.

{\em Contextual mutants in mutation testing.} 
While we introduce contextual mutations in the domain of neural bug detection, the idea
of biasing the mutation on the code context has been explored in mutation testing~\cite{just2017inferring, allamanis2016tailored}.
Just et al.~\cite{just2017inferring} inferred the mutant utility based on the AST parents and children
surrounding the mutation location leading to a context-based mutant selection. 
While our method does not take the AST into account, our mutator can exploit natural
hints in program code to produce mutants in code. 
Allamanis et al.~\cite{allamanis2016tailored} showed that this type of information can
improve the effectiveness of mutant generation by inferring mutant utility via an ngram language model.
In contrast to our work, the authors select mutants that are most unlikely in the mutation context. While
this is a promising strategy to find corner cases in test suites, we believe that
neural bug detectors need to learn from more realistic mutants to achieve a good detection performance.

{\em Training data generation for language processing.}
With a similarly high demand in training data and lack of labeled data in natural 
language processing, much effort has been spent to pre-train neural language representations on massive corpora of unlabeled text~\cite{devlin2019bert, clark2020electra, raffel2019exploring} 
that can be fine-tuned to a task with fewer labeled  examples. Devlin et al.~\cite{devlin2019bert} were, for example, the first
to report performance improvements across several NLP benchmarks by pre-training with the masked
language model. 
While our technique is inspired by pre-training
techniques \cite{clark2020electra}, our goal is not to provide a general programming language representation,
but to generate more realistic buggy programs. As a consequence, all our models
are trained for specific bug targets, while their internally learned representation
might not generalize well to other bug types. Surprisingly, our experiments
showed that "fine-tuning" our models to a specific mutator improves
results on synthetic benchmarks (although this is not always reflected by real world performance).
 

%% file: conclusion.tex
\section{Conclusion}
In this paper, we propose a novel contextual mutation operator that learns to inject bugs fitting to the 
mutation context. For this purpose, our mutator samples token replacements from a continuously
trained masked language model.
Our evaluation shows that using a learning-based mutator increases the effectiveness
on both synthetic and real world benchmarks. In addition, our experiments show that a careful choice 
of the employed mutator and a dynamic bug injection strategy have in general
a positive effect on the performance of a bug detector. 
Based on this result, we recommend that future practitioners spend significant
effort in the careful design of bug injection strategies.

In the future, we intend to apply our contextual mutator for the construction
of better synthetic benchmarks and training of more advanced future bug detectors.

%% file: main.bbl
\begin{thebibliography}{10}
\providecommand{\url}[1]{#1}
\csname url@samestyle\endcsname
\providecommand{\newblock}{\relax}
\providecommand{\bibinfo}[2]{#2}
\providecommand{\BIBentrySTDinterwordspacing}{\spaceskip=0pt\relax}
\providecommand{\BIBentryALTinterwordstretchfactor}{4}
\providecommand{\BIBentryALTinterwordspacing}{\spaceskip=\fontdimen2\font plus
\BIBentryALTinterwordstretchfactor\fontdimen3\font minus
  \fontdimen4\font\relax}
\providecommand{\BIBforeignlanguage}[2]{{%
\expandafter\ifx\csname l@#1\endcsname\relax
\typeout{** WARNING: IEEEtran.bst: No hyphenation pattern has been}%
\typeout{** loaded for the language `#1'. Using the pattern for}%
\typeout{** the default language instead.}%
\else
\language=\csname l@#1\endcsname
\fi
#2}}
\providecommand{\BIBdecl}{\relax}
\BIBdecl

\bibitem{habib2019neural}
A.~Habib and M.~Pradel, ``Neural bug finding: A study of opportunities and
  challenges,'' \emph{arXiv preprint arXiv:1906.00307}, 2019.

\bibitem{li2019improving}
Y.~Li, S.~Wang, T.~N. Nguyen, and S.~Van~Nguyen, ``Improving bug detection via
  context-based code representation learning and attention-based neural
  networks,'' \emph{Proceedings of the ACM on Programming Languages}, vol.~3,
  no. OOPSLA, pp. 1--30, 2019.

\bibitem{briem2019using}
J.~A. Briem, J.~Smit, H.~Sellik, and P.~Rapoport, ``Using distributed
  representation of code for bug detection,'' \emph{arXiv preprint
  arXiv:1911.12863}, 2019.

\bibitem{allamanis18graphs}
\BIBentryALTinterwordspacing
M.~Allamanis, M.~Brockschmidt, and M.~Khademi, ``Learning to represent programs
  with graphs,'' in \emph{6th International Conference on Learning
  Representations, {ICLR} 2018, Vancouver, BC, Canada, April 30 - May 3, 2018,
  Conference Track Proceedings}.\hskip 1em plus 0.5em minus 0.4em\relax
  OpenReview.net, 2018. [Online]. Available:
  \url{https://openreview.net/forum?id=BJOFETxR-}
\BIBentrySTDinterwordspacing

\bibitem{vasic2019joint}
\BIBentryALTinterwordspacing
M.~Vasic, A.~Kanade, P.~Maniatis, D.~Bieber, and R.~Singh, ``Neural program
  repair by jointly learning to localize and repair,'' in \emph{7th
  International Conference on Learning Representations, {ICLR} 2019, New
  Orleans, LA, USA, May 6-9, 2019}.\hskip 1em plus 0.5em minus 0.4em\relax
  OpenReview.net, 2019. [Online]. Available:
  \url{https://openreview.net/forum?id=ByloJ20qtm}
\BIBentrySTDinterwordspacing

\bibitem{hellendoorn2020great}
\BIBentryALTinterwordspacing
V.~J. Hellendoorn, C.~Sutton, R.~Singh, P.~Maniatis, and D.~Bieber, ``Global
  relational models of source code,'' in \emph{8th International Conference on
  Learning Representations, {ICLR} 2020, Addis Ababa, Ethiopia, April 26-30,
  2020}.\hskip 1em plus 0.5em minus 0.4em\relax OpenReview.net, 2020. [Online].
  Available: \url{https://openreview.net/forum?id=B1lnbRNtwr}
\BIBentrySTDinterwordspacing

\bibitem{pradel2018deepbugs}
M.~Pradel and K.~Sen, ``Deepbugs: A learning approach to name-based bug
  detection,'' \emph{Proceedings of the ACM on Programming Languages}, vol.~2,
  no. OOPSLA, pp. 1--25, 2018.

\bibitem{Ray2016Natural}
\BIBentryALTinterwordspacing
B.~Ray, V.~Hellendoorn, S.~Godhane, Z.~Tu, A.~Bacchelli, and P.~Devanbu, ``On
  the "naturalness" of buggy code,'' in \emph{Proceedings of the 38th
  International Conference on Software Engineering}, ser. ICSE '16.\hskip 1em
  plus 0.5em minus 0.4em\relax New York, NY, USA: Association for Computing
  Machinery, 2016, p. 428–439. [Online]. Available:
  \url{https://doi.org/10.1145/2884781.2884848}
\BIBentrySTDinterwordspacing

\bibitem{karampatsis2020scelmo}
R.-M. Karampatsis and C.~Sutton, ``Scelmo: Source code embeddings from language
  models,'' \emph{arXiv preprint arXiv:2004.13214}, 2020.

\bibitem{Smith2009Augment}
\BIBentryALTinterwordspacing
B.~H. Smith and L.~Williams, ``On guiding the augmentation of an automated test
  suite via mutation analysis,'' \emph{Empirical Softw. Engg.}, vol.~14, no.~3,
  p. 341–369, Jun. 2009. [Online]. Available:
  \url{https://doi.org/10.1007/s10664-008-9083-7}
\BIBentrySTDinterwordspacing

\bibitem{just2014major}
R.~Just, ``The major mutation framework: Efficient and scalable mutation
  analysis for java,'' in \emph{Proceedings of the 2014 international symposium
  on software testing and analysis}, 2014, pp. 433--436.

\bibitem{namin2008sufficient}
\BIBentryALTinterwordspacing
A.~Siami~Namin, J.~H. Andrews, and D.~J. Murdoch, ``Sufficient mutation
  operators for measuring test effectiveness,'' in \emph{Proceedings of the
  30th International Conference on Software Engineering}, ser. ICSE '08.\hskip
  1em plus 0.5em minus 0.4em\relax New York, NY, USA: Association for Computing
  Machinery, 2008, p. 351–360. [Online]. Available:
  \url{https://doi.org/10.1145/1368088.1368136}
\BIBentrySTDinterwordspacing

\bibitem{allamanis2016tailored}
\BIBentryALTinterwordspacing
M.~Allamanis, E.~T. Barr, R.~Just, and C.~Sutton, ``Tailored mutants fit bugs
  better,'' \emph{CoRR}, vol. abs/1611.02516, 2016. [Online]. Available:
  \url{http://arxiv.org/abs/1611.02516}
\BIBentrySTDinterwordspacing

\bibitem{just2014realbugs}
\BIBentryALTinterwordspacing
R.~Just, D.~Jalali, L.~Inozemtseva, M.~D. Ernst, R.~Holmes, and G.~Fraser,
  ``Are mutants a valid substitute for real faults in software testing?'' in
  \emph{Proceedings of the 22nd ACM SIGSOFT International Symposium on
  Foundations of Software Engineering}, ser. FSE 2014.\hskip 1em plus 0.5em
  minus 0.4em\relax New York, NY, USA: Association for Computing Machinery,
  2014, p. 654–665. [Online]. Available:
  \url{https://doi.org/10.1145/2635868.2635929}
\BIBentrySTDinterwordspacing

\bibitem{daran96realbugs}
\BIBentryALTinterwordspacing
M.~Daran and P.~Th\'{e}venod-Fosse, ``Software error analysis: A real case
  study involving real faults and mutations,'' in \emph{Proceedings of the 1996
  ACM SIGSOFT International Symposium on Software Testing and Analysis}, ser.
  ISSTA '96.\hskip 1em plus 0.5em minus 0.4em\relax New York, NY, USA:
  Association for Computing Machinery, 1996, p. 158–171. [Online]. Available:
  \url{https://doi.org/10.1145/229000.226313}
\BIBentrySTDinterwordspacing

\bibitem{papadakis2018mutation}
M.~Papadakis, D.~Shin, S.~Yoo, and D.-H. Bae, ``Are mutation scores correlated
  with real fault detection? a large scale empirical study on the relationship
  between mutants and real faults,'' in \emph{2018 IEEE/ACM 40th International
  Conference on Software Engineering (ICSE)}.\hskip 1em plus 0.5em minus
  0.4em\relax IEEE, 2018, pp. 537--548.

\bibitem{just2012using}
R.~Just, G.~M. Kapfhammer, and F.~Schweiggert, ``Using non-redundant mutation
  operators and test suite prioritization to achieve efficient and scalable
  mutation analysis,'' in \emph{2012 IEEE 23rd International Symposium on
  Software Reliability Engineering}.\hskip 1em plus 0.5em minus 0.4em\relax
  IEEE, 2012, pp. 11--20.

\bibitem{just2017inferring}
R.~Just, B.~Kurtz, and P.~Ammann, ``Inferring mutant utility from program
  context,'' in \emph{Proceedings of the International Symposium on Software
  Testing and Analysis (ISSTA)}, Santa Barbara, CA, USA, July~10--12 2017, pp.
  284--294.

\bibitem{devlin2019bert}
\BIBentryALTinterwordspacing
J.~Devlin, M.-W. Chang, K.~Lee, and K.~Toutanova, ``{BERT}: Pre-training of
  deep bidirectional transformers for language understanding,'' in
  \emph{Proceedings of the 2019 Conference of the North {A}merican Chapter of
  the Association for Computational Linguistics: Human Language Technologies,
  Volume 1 (Long and Short Papers)}.\hskip 1em plus 0.5em minus 0.4em\relax
  Minneapolis, Minnesota: Association for Computational Linguistics, Jun. 2019,
  pp. 4171--4186. [Online]. Available:
  \url{https://www.aclweb.org/anthology/N19-1423}
\BIBentrySTDinterwordspacing

\bibitem{clark2020electra}
\BIBentryALTinterwordspacing
K.~Clark, M.~Luong, Q.~V. Le, and C.~D. Manning, ``{ELECTRA:} pre-training text
  encoders as discriminators rather than generators,'' in \emph{8th
  International Conference on Learning Representations, {ICLR} 2020, Addis
  Ababa, Ethiopia, April 26-30, 2020}.\hskip 1em plus 0.5em minus 0.4em\relax
  OpenReview.net, 2020. [Online]. Available:
  \url{https://openreview.net/forum?id=r1xMH1BtvB}
\BIBentrySTDinterwordspacing

\bibitem{allamanis2018survey}
\BIBentryALTinterwordspacing
M.~Allamanis, E.~T. Barr, P.~Devanbu, and C.~Sutton, ``A survey of machine
  learning for big code and naturalness,'' \emph{ACM Comput. Surv.}, vol.~51,
  no.~4, Jul. 2018. [Online]. Available: \url{https://doi.org/10.1145/3212695}
\BIBentrySTDinterwordspacing

\bibitem{hindle2012natural}
A.~Hindle, E.~T. Barr, Z.~Su, M.~Gabel, and P.~Devanbu, ``On the naturalness of
  software,'' in \emph{Proceedings of the 34th International Conference on
  Software Engineering}, ser. ICSE '12.\hskip 1em plus 0.5em minus 0.4em\relax
  IEEE Press, 2012, p. 837–847.

\bibitem{alon2018code2seq}
U.~Alon, S.~Brody, O.~Levy, and E.~Yahav, ``code2seq: Generating sequences from
  structured representations of code,'' \emph{arXiv preprint arXiv:1808.01400},
  2018.

\bibitem{iyer2016summarizing}
\BIBentryALTinterwordspacing
S.~Iyer, I.~Konstas, A.~Cheung, and L.~Zettlemoyer, ``Summarizing source code
  using a neural attention model,'' in \emph{Proceedings of the 54th Annual
  Meeting of the Association for Computational Linguistics (Volume 1: Long
  Papers)}.\hskip 1em plus 0.5em minus 0.4em\relax Berlin, Germany: Association
  for Computational Linguistics, Aug. 2016, pp. 2073--2083. [Online].
  Available: \url{https://www.aclweb.org/anthology/P16-1195}
\BIBentrySTDinterwordspacing

\bibitem{ahmad2020transformer}
W.~U. Ahmad, S.~Chakraborty, B.~Ray, and K.-W. Chang, ``A transformer-based
  approach for source code summarization,'' \emph{arXiv preprint
  arXiv:2005.00653}, 2020.

\bibitem{lachaux2020unsupervised}
M.-A. Lachaux, B.~Roziere, L.~Chanussot, and G.~Lample, ``Unsupervised
  translation of programming languages,'' \emph{arXiv preprint
  arXiv:2006.03511}, 2020.

\bibitem{arnar2020offside}
\BIBentryALTinterwordspacing
J.~A. Briem, J.~Smit, H.~Sellik, P.~Rapoport, G.~Gousios, and M.~Aniche,
  ``Offside: Learning to identify mistakes in boundary conditions,'' in
  \emph{Proceedings of the IEEE/ACM 42nd International Conference on Software
  Engineering Workshops}, ser. ICSEW'20.\hskip 1em plus 0.5em minus 0.4em\relax
  New York, NY, USA: Association for Computing Machinery, 2020, p. 203–208.
  [Online]. Available: \url{https://doi.org/10.1145/3387940.3391464}
\BIBentrySTDinterwordspacing

\bibitem{karampatsis20bigcode}
\BIBentryALTinterwordspacing
R.-M. Karampatsis, H.~Babii, R.~Robbes, C.~Sutton, and A.~Janes, ``Big code !=
  big vocabulary: Open-vocabulary models for source code,'' in
  \emph{Proceedings of the ACM/IEEE 42nd International Conference on Software
  Engineering}, ser. ICSE '20.\hskip 1em plus 0.5em minus 0.4em\relax New York,
  NY, USA: Association for Computing Machinery, 2020, p. 1073–1085. [Online].
  Available: \url{https://doi.org/10.1145/3377811.3380342}
\BIBentrySTDinterwordspacing

\bibitem{zhaopeng2014localness}
\BIBentryALTinterwordspacing
Z.~Tu, Z.~Su, and P.~Devanbu, ``On the localness of software,'' in
  \emph{Proceedings of the 22nd ACM SIGSOFT International Symposium on
  Foundations of Software Engineering}, ser. FSE 2014.\hskip 1em plus 0.5em
  minus 0.4em\relax New York, NY, USA: Association for Computing Machinery,
  2014, p. 269–280. [Online]. Available:
  \url{https://doi.org/10.1145/2635868.2635875}
\BIBentrySTDinterwordspacing

\bibitem{hellendoorn2017deep}
V.~J. Hellendoorn and P.~Devanbu, ``Are deep neural networks the best choice
  for modeling source code?'' in \emph{Proceedings of the 2017 11th Joint
  Meeting on Foundations of Software Engineering}, 2017, pp. 763--773.

\bibitem{karampatsis2020sstubs}
\BIBentryALTinterwordspacing
R.-M. Karampatsis and C.~Sutton, ``How often do single-statement bugs occur?
  the manysstubs4j dataset,'' in \emph{Proceedings of the 17th International
  Conference on Mining Software Repositories}, ser. MSR '20.\hskip 1em plus
  0.5em minus 0.4em\relax New York, NY, USA: Association for Computing
  Machinery, 2020, p. 573–577. [Online]. Available:
  \url{https://doi.org/10.1145/3379597.3387491}
\BIBentrySTDinterwordspacing

\bibitem{just2014d4j}
\BIBentryALTinterwordspacing
R.~Just, D.~Jalali, and M.~D. Ernst, ``Defects4j: A database of existing faults
  to enable controlled testing studies for java programs,'' in
  \emph{Proceedings of the 2014 International Symposium on Software Testing and
  Analysis}, ser. ISSTA 2014.\hskip 1em plus 0.5em minus 0.4em\relax New York,
  NY, USA: Association for Computing Machinery, 2014, p. 437–440. [Online].
  Available: \url{https://doi.org/10.1145/2610384.2628055}
\BIBentrySTDinterwordspacing

\bibitem{spotbugs}
\BIBentryALTinterwordspacing
(2021) Spot bugs. [Online]. Available: \url{https://spotbugs.github.io/}
\BIBentrySTDinterwordspacing

\bibitem{sennrich2016neural}
\BIBentryALTinterwordspacing
R.~Sennrich, B.~Haddow, and A.~Birch, ``Neural machine translation of rare
  words with subword units,'' in \emph{Proceedings of the 54th Annual Meeting
  of the Association for Computational Linguistics (Volume 1: Long
  Papers)}.\hskip 1em plus 0.5em minus 0.4em\relax Berlin, Germany: Association
  for Computational Linguistics, Aug. 2016, pp. 1715--1725. [Online].
  Available: \url{https://www.aclweb.org/anthology/P16-1162}
\BIBentrySTDinterwordspacing

\bibitem{shaw2018relative}
\BIBentryALTinterwordspacing
P.~Shaw, J.~Uszkoreit, and A.~Vaswani, ``Self-attention with relative position
  representations,'' in \emph{Proceedings of the 2018 Conference of the North
  American Chapter of the Association for Computational Linguistics: Human
  Language Technologies, NAACL-HLT, New Orleans, Louisiana, USA, June 1-6,
  2018, Volume 2 (Short Papers)}, M.~A. Walker, H.~Ji, and A.~Stent, Eds.\hskip
  1em plus 0.5em minus 0.4em\relax Association for Computational Linguistics,
  2018, pp. 464--468. [Online]. Available:
  \url{https://doi.org/10.18653/v1/n18-2074}
\BIBentrySTDinterwordspacing

\bibitem{husain2019codesearch}
\BIBentryALTinterwordspacing
H.~Husain, H.~Wu, T.~Gazit, M.~Allamanis, and M.~Brockschmidt, ``Codesearchnet
  challenge: Evaluating the state of semantic code search,'' \emph{CoRR}, vol.
  abs/1909.09436, 2019. [Online]. Available:
  \url{http://arxiv.org/abs/1909.09436}
\BIBentrySTDinterwordspacing

\bibitem{ray2016model}
\BIBentryALTinterwordspacing
V.~Raychev, P.~Bielik, and M.~T. Vechev, ``Probabilistic model for code with
  decision trees,'' in \emph{Proceedings of the 2016 {ACM} {SIGPLAN}
  International Conference on Object-Oriented Programming, Systems, Languages,
  and Applications, {OOPSLA} 2016, part of {SPLASH} 2016, Amsterdam, The
  Netherlands, October 30 - November 4, 2016}, E.~Visser and Y.~Smaragdakis,
  Eds.\hskip 1em plus 0.5em minus 0.4em\relax {ACM}, 2016, pp. 731--747.
  [Online]. Available: \url{https://doi.org/10.1145/2983990.2984041}
\BIBentrySTDinterwordspacing

\bibitem{hochreiter1997long}
S.~Hochreiter and J.~Schmidhuber, ``Long short-term memory,'' \emph{Neural
  computation}, vol.~9, no.~8, pp. 1735--1780, 1997.

\bibitem{kingma2015adam}
\BIBentryALTinterwordspacing
D.~P. Kingma and J.~Ba, ``Adam: {A} method for stochastic optimization,'' in
  \emph{3rd International Conference on Learning Representations, {ICLR} 2015,
  San Diego, CA, USA, May 7-9, 2015, Conference Track Proceedings}, Y.~Bengio
  and Y.~LeCun, Eds., 2015. [Online]. Available:
  \url{http://arxiv.org/abs/1412.6980}
\BIBentrySTDinterwordspacing

\bibitem{dinella2020hoppity}
E.~Dinella, H.~Dai, Z.~Li, M.~Naik, L.~Song, and K.~Wang, ``Hoppity: Learning
  graph transformations to detect and fix bugs in programs,'' in
  \emph{International Conference on Learning Representations (ICLR)}, 2020.

\bibitem{raffel2019exploring}
C.~Raffel, N.~Shazeer, A.~Roberts, K.~Lee, S.~Narang, M.~Matena, Y.~Zhou,
  W.~Li, and P.~J. Liu, ``Exploring the limits of transfer learning with a
  unified text-to-text transformer,'' \emph{arXiv preprint arXiv:1910.10683},
  2019.

\end{thebibliography}
